\documentclass[twocolumn,showpacs,preprintnumbers,amsmath,amssymb,superscriptaddress]{revtex4}

\usepackage{graphicx}
\usepackage{dcolumn}
\usepackage{bm}

\begin{document}

\title{Absorption and photoluminescence spectroscopy on a single self-assembled charge-tunable quantum dot}

\author{S. Seidl}
\affiliation{Center for NanoScience and Sektion Physik, Ludwig-Maximilians-Universit\"{a}t, Geschwister-Scholl-Platz 1, 80539 M\"{u}nchen, Germany}
\affiliation{School of Engineering and Physical Sciences, Heriot-Watt University, Edinburgh EH14 4AS, UK}
\email{S.Seidl@hw.ac.uk}

\author{M. Kroner}
\affiliation{Center for NanoScience and Sektion Physik, Ludwig-Maximilians-Universit\"{a}t, Geschwister-Scholl-Platz 1, 80539 M\"{u}nchen, Germany}

\author{P. A. Dalgarno}
\affiliation{School of Engineering and Physical Sciences, Heriot-Watt University, Edinburgh EH14 4AS, UK}

\author{A. H\"{o}gele}
\affiliation{Center for NanoScience and Sektion Physik, Ludwig-Maximilians-Universit\"{a}t, Geschwister-Scholl-Platz 1, 80539 M\"{u}nchen, Germany}

\author{J. M. Smith}
\affiliation{School of Engineering and Physical Sciences, Heriot-Watt University, Edinburgh EH14 4AS, UK}

\author{M. Ediger}
\affiliation{School of Engineering and Physical Sciences, Heriot-Watt University, Edinburgh EH14 4AS, UK}

\author{B. D. Gerardot}
\affiliation{School of Engineering and Physical Sciences, Heriot-Watt University, Edinburgh EH14 4AS, UK}
\affiliation{Materials Department, University of California, Santa Barbara, California 93106, USA}

\author{J. M. Garcia}
\affiliation{Materials Department, University of California, Santa Barbara, California 93106, USA}
\affiliation{Instituto de  Microelectr\'{o}nica de Madrid, CNM (CSIC), Isaac Newton 8, PTM, 28760 Tres Cantos, Madrid, Spain}

\author{P. M. Petroff}
\affiliation{Materials Department, University of California, Santa Barbara, California 93106, USA}

\author{K. Karrai}
\affiliation{Center for NanoScience and Sektion Physik, Ludwig-Maximilians-Universit\"{a}t, Geschwister-Scholl-Platz 1, 80539 M\"{u}nchen, Germany}

\author{R. J. Warburton}
\affiliation{School of Engineering and Physical Sciences, Heriot-Watt University, Edinburgh EH14 4AS, UK}

\date{\today}

\begin{abstract}
We have performed detailed photoluminescence (PL) and absorption spectroscopy on the same single self-assembled quantum dot in a charge-tunable device. The transition from neutral to charged exciton in the PL occurs at a more negative voltage than the corresponding transition in absorption. We have developed a model of the Coulomb blockade to account for this observation. At large negative bias, the absorption broadens as a result of electron and hole tunneling. We observe resonant features in this regime whenever the quantum dot hole level is resonant with two-dimensional hole states located at the capping layer-blocking barrier interface in our structure. 
\end{abstract}

\maketitle

The strong quantization in self-assembled quantum dots (QDs) makes this system of great interest for quantum optics and quantum information processing \cite{ImamogluPRL1999,BensonPRL2000,MichlerScience2000}. It has been shown that single self assembled QDs can be controllably charged with single electrons \cite{WarburtonNature2000}. The energy shifts, fine-structure splittings \cite{WarburtonPRL2003} and the behavior in magnetic field \cite{KarraiNature2004} of differently charged QD states have been probed by photoluminescence (PL) spectroscopy. However, PL spectroscopy has a major drawback as it invariably involves excitation into a state well above the ground state. The subsequent relaxation step is incoherent, generally causing phase and spin information to be lost. This disadvantage is completely eliminated by resonant excitation, detecting exciton creation with absorption spectroscopy. However, absorption spectroscopy gives a small contrast in the transmitted light intensity and it therefore remains a challenging experiment, especially on a single self-assembled QDs where the oscillator strength is small \cite{WarburtonPRL1997}. Nevertheless, absorption spectroscopy has now been achieved on a single QD with good signal to noise \cite{BonadeoAPL99,AlenAPL2003,HoegelePRL2004}. An important and hitherto unanswered question is the exact relationship between PL and absorption spectroscopy.

We present here both PL and absorption measurements carried out on the same QD. Naively, one would assume that both the PL and absorption processes involve the same levels, the PL a deexcitation process, absorption an excitation process. This idea turns out to be too simplistic in our case. By embedding the QD in a field-effect structure and working in the Coulomb blockade regime, we uncover a striking difference between PL and absorption. In a certain gate voltage regime, the final state of the absorption transition is not the initial state of the PL transition, and the final state of the PL is not the initial state of the absorption. We show that the origin of this remarkable effect lies in the difference between the electron-electron and electron-hole on-site Coulomb energies. We show further that we can control the homogenous linewidth. At large electric field, both electron and hole tunnel out of the QD after resonant excitation such that the linewidth represents the tunneling rate. The electron tunneling rate increases monotonically with increasing electric field whereas the hole tunneling rate oscillates as resonances are established with two-dimensional hole states.

\begin{figure}
\includegraphics{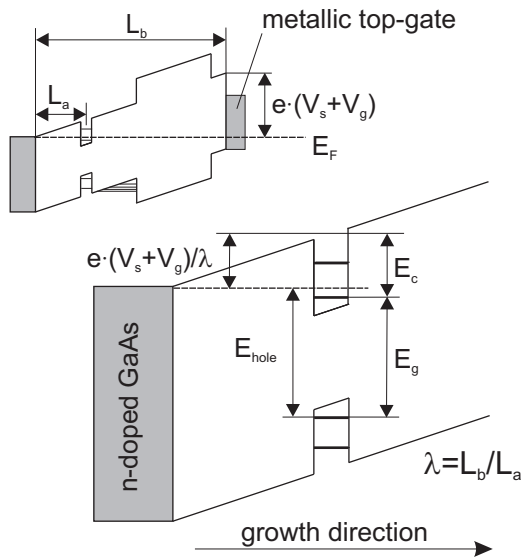}
\caption{\label{fig:structure} Schematic sketches of the heterostructure showing the conduction and valence band edges along the growth direction. A voltage $V_{g}$ is applied to the gate electrode. An electric field results from both the applied field and the Schottky voltage $V_{s}$. The QDs are separated from the n-doped GaAs by a tunnel barrier, 25 nm of undoped GaAs, and from the blocking barrier by 30 nm of undoped GaAs. The lever arm of the device is $\lambda=7$. The lower diagram labels the energies used in the electrostatic calculations of the device.}
\end{figure}

The semiconductor heterostructure consists of a layer of self-assembled QDs embedded in an n-type FET-structure as shown in Fig.\ \ref{fig:structure}. By applying an increasing reverse bias voltage ($V_g$) between the metallic top gate and the back gate, the QD energy levels are raised relative to the Fermi energy. This causes a pronounced Coulomb blockade both in the electron occupation \cite{WarburtonPRB1998} and, under weak optical excitation, in the exciton charge \cite{WarburtonNature2000}. A blocking barrier in the heterostructure prevents electron tunneling from the top gate. PL was carried out with non-resonant excitation at 830 nm wavelength, the PL detected with a grating spectrometer and multi-channel Si detector with spectral resolution 100 $\mu$eV. The absorption spectroscopy relied on a narrow band laser. The power was kept below 10 nW in order to avoid saturating the resonance, and the laser light was detected beneath the sample with an situ Ge p-i-n diode. We swept the QD energy through the constant laser energy by exploiting the small vertical Stark effect \cite{AlenAPL2003} while modulating the gate voltage with a 100 mV peak to peak square wave in order to reject random noise in the detector current with a lock-in amplifier. All experiments were performed at 4.2 K with a confocal microscope with 1 $\mu$m resolution.

Fig.\ 2(a) shows a PL spectrum of a negatively charged exciton, X$^{1-}$. The linewidth corresponds to our spectral resolution, showing that the real QD line shape is not accessible with this experimental setup. This is typical for grating spectrometers which have linewidths of a few tens of $\mu$eV. Fig.\ 2(b) shows an absorption spectrum measured on the same QD. In this case, the experimental resolution is 0.01 $\mu$eV such that the experiment accesses the proper QD line shape. In Fig.\ 2(b), we observe a linewidth of 2.3 $\mu$eV. The limit of radiative lifetime broadening corresponds to a linewidth of about 1 $\mu$eV for this QD. It is very likely that the largest contribution to the additional broadening is caused by temporal fluctuations of the resonance energy during the integration time (1 s in Fig.\ 2(b)) \cite{HoegelePRL2004}.

\begin{figure}
\includegraphics{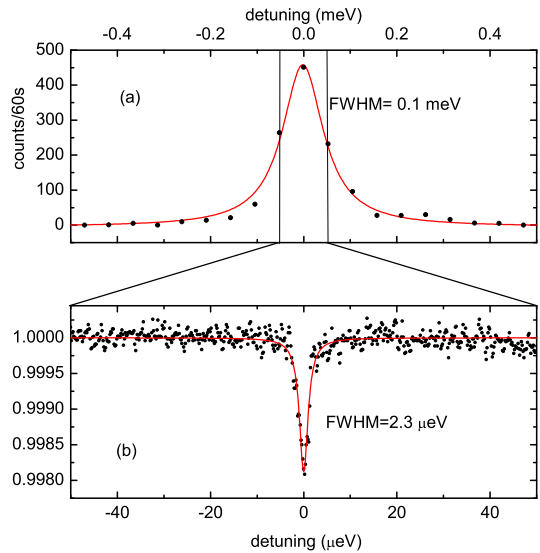}
\caption{(a) PL spectrum of an X$^{1-}$ exciton. The linewidth, 100 $\mu$eV, is determined by the resolution of the setup and not by the QD itself. (b) Absorption spectrum on the same QD as (a). The resonance corresponds to excitation of the X$^{1-}$. The resolution is not setup limited and the resonance exhibits a linewidth of 2.3 $\mu$eV. Both spectra were measured at the same gate voltage and at a temperature of 4.2 K. The solid lines are Lorentz curves fitted to the data points.}
\end{figure}

The higher spectral resolution is not the only advantage of absorption spectroscopy over PL for our sample structure. Fig.\ 3 shows the resonant gate voltage for constant energy for both measurement techniques as a function of excitation intensity. We find that the resonant gate voltage has a strong intensity dependence in the case of PL but no measurable intensity dependence in the case of absorption.  This makes the interpretation of the charging voltages in PL problematic as the resonant gate voltage is intensity dependent right down to extremely low excitation powers as shown in Fig.\ 3. The explanation for the intensity-dependent resonant gate voltages in the PL is that some of the holes excited by the PL pump laser become trapped at the interface between the capping layer and the blocking barrier (Fig.\ \ref{fig:structure}), creating a space charge and hence an internal electric field, opposite in polarity to the applied bias \cite{SmithAPL2003}. In our case, we excite the PL non-resonantly. The photo-excited electrons have a small density compared to the density of electrons in the back contact such that the hole plays the more important role in exciting the PL. Some of the holes relax into the dots but others relax from the wetting layer to the capping layer-blocking barrier interface, creating a large space charge. Resonant excitation would decrease the space charge effects but probably not eliminate them and in fact whenever an electric field is applied without a current flow, an insulating layer is needed and excitation of the PL will build up a space charge. In the absorption experiment, we address just one QD with laser light resonant with the ground state exciton such that no space charge is established and the resonant voltage is independent of the pump power.

\begin{figure}
\includegraphics{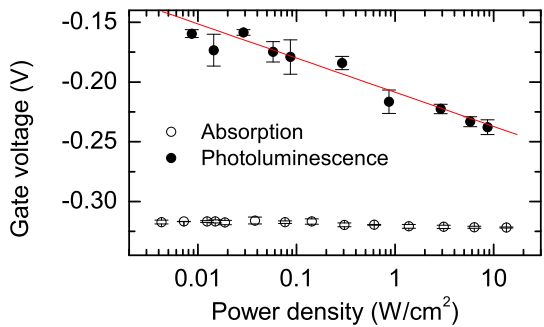}
\caption{Resonant gate voltage against pump power density for PL and absorption spectroscopy, in both cases keeping the wavelength constant. The data were taken from two different QDs at 4.2 K and represent typical behavior for the QDs embedded in the heterostructure of Fig.\ 1. The line describing the PL data is a guide to the eye.}
\end{figure}

The exciton charge changes abruptly from X$^{0}$ to X$^{1-}$ on changing the gate voltage in both absorption and PL experiments, as shown in Fig.\ 4(a). In both cases, this is a consequence of the large Coulomb interactions and fast tunneling in our heterostructure. Once a quantum dot captures a hole, electrons tunnel into the dot from the back contact to form the configuration which has the lowest energy at that particular voltage. Fig.\ 4(a), a plot of the resonance energy as a function of gate voltage, shows that there is a clear red-shift on going from X$^{0}$ to X$^{1-}$. This arises through the Coulomb interaction and can be interpreted as a band gap renormalization in the single electron limit \cite{WarburtonNature2000}. 

In the case of absorption, the X$^{0}$ disappears and is replaced by the X$^{1-}$ at a particular voltage, $V_g=V_3$. The X$^{1-}$ exists at more positive voltages, disappearing at $V_g=V_4$. The X$^0$ signal persists to voltages considerably less than $V_3$ but the linewidth shows a strong voltage dependence, Figs.\ 5 and 6. The X$^{0}$ absorption linewdith is small, 3 $\mu$eV, between voltages $V_1$ and $V_3$ but at $V_1$ the linewidth starts to increase, and the linewidth increases further as $V_g$ is made more negative with oscillatory features as shown in Fig.\ 6. The absorption linewidth increases to more than 100 $\mu$eV at the largest electric fields where the contrast at resonance is very small, as shown in Fig.\ 5, making its detection challenging.

\begin{figure}
\includegraphics{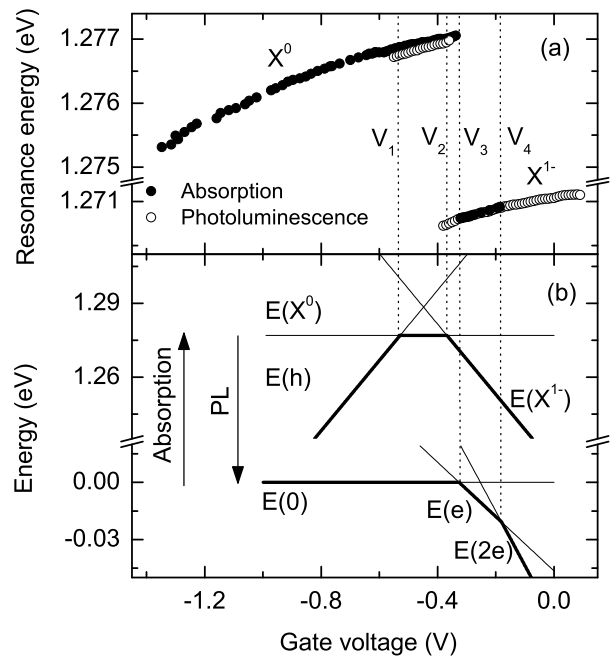}
\caption{Behavior of (a) the PL and absorption energies as a function of gate voltage, $V_g$. The PL energies are corrected for space charge effects as described in the text. (b) Energy of the 0, e, 2e; h, X$^{0}$, X$^{1-}$ states as a function of $V_g$. The bold lines show the energetically favored states with (upper part) and without (lower part) a hole.}
\end{figure}

The PL shows a different behavior. To make a detailed comparison with the absorption data it is necessary to correct the PL data for the space charge effects. Gauss' law reveals that the space charge at the interface between the capping layer and the blocking barrier shifts the resonant voltages rigidly without changing the lever arm. The experimental data support this view. We find for instance that the X$^0$ plateau moves to lower applied biases as the excitation intensity is increased as a consequence of the increased space charge but the voltage extent of the X$^0$ plateau remains the same. This result also shows that the space charge density is independent of bias over the voltage range of interest in this experiment. We can estimate the correction in the resonant voltages from Fig.\ 3 due to the space charge, measuring at one wavelength the difference in the resonant voltage at the power used in the measurement and the resonant voltage in the limit of extremely low power. However, the sensitivity of the resonant voltage to pump power even at very low powers introduces an uncertainty of more than 0.05 V in the correction for space charge. As we show however, the X$^{1-}$ PL linewidth provides a much more accurate measure of this correction, facilitating a detailed comparison between PL and absorption. Fig.\ 4(a) shows both the PL energies corrected for space charge and the absorption energies versus gate voltage.

The X$^{0}$ PL first appears at $V_g=V_1$, the voltage at which the X$^0$ absorption starts to broaden. For $V_g<V_1$, there is no X$^0$ PL whatsoever but, as shown in Fig.\ 4(a), there is a strong X$^0$ absorption resonance. At $V_g=V_2$, the X$^{0}$ PL is replaced by the X$^{1-}$ PL. The significant result is that $V_2$ is less than $V_3$ by 46 mV. On the more positive voltage side, while the X$^{1-}$ absorption disappears at $V_4$, the X$^{1-}$ PL extends to much larger voltages.

We explain the behavior of both absorption and PL with a unifying model of the Coulomb blockade. The energies of the charges in the QD are perturbed by their mutual Coulomb interactions and the electrostatic potential. This potential is the sum of the gate voltage $V_{g}$ and the Schottky voltage $V_{s}$ divided by the lever arm $\lambda$, which is 7 for our structure \cite{WarburtonPRB1998}. Therefore a change in energy of 1 meV corresponds to a 7 mV gate voltage difference. We treat the Coulomb interactions as perturbations to the quantized level structure which is a good approximation for the strongly-confined QDs used here \cite{SchulhauserPRB2002}. We ignore the Stark effect as it is small compared to the Coulomb energies but we do include the interactions of charges with the mirror charges in the back contact \cite{WuEJP2000}. Fig.\ \ref{fig:structure} defines the important energies in the model. The energy of an electron at the Fermi energy in the back contact is taken to be zero. To charge the QD with a single electron, an electron must overcome the electrostatic barrier, the only gate voltage-dependent parameter in the model, but the electron gains the confinement energy of the QD. Charging the QD with an additional electron involves the same terms but also an on-site Coulomb energy. To add a hole, we consider an electron being transferred from the highest energy valence level to the Fermi energy. The hole has a negative on-site Coulomb energy with any confined electrons in the QD. Table \ref{tab:table1} lists the energies of all the states of interest in this experiment within these approximations. The energies of the different states have a linear dependence on the gate voltage and are plotted in Fig.\ 4(b). The states in the upper (lower) part of the figure correspond to a QD which is occupied (unoccupied) by a hole. 

As a function of gate voltage, the nature of the state with lowest energy changes when the energy of one state crosses below that of another. Without a hole, the empty QD state $E(0)$ is preferred for very negative gate voltages. At $V_{g}=V_{3}$, $E(0)$ becomes equal to the energy of the state containing a single electron in the QD, $E(e)$, such that an electron is trapped in the QD if the voltage is increased further. A second electron cannot enter the QD until the gate voltage is increased to $V_{4}$ because of the Coulomb blockade. These voltages are analogous to the current peaks in transport measurements on a single QD \cite{KouwenhovenRPP2001}. Similarly, for the states containing a hole in Fig.\ 4(b), again the nature of the ground state changes as a function of $V_{g}$, in this case from the hole only state to X$^{0}$ at $V_{1}$ and then from X$^0$ to X$^{1-}$ at $V_{2}$. The change in the nature of the ground state as a function of $V_{g}$ corresponds exactly to the excitonic Coulomb blockade \cite{WarburtonNature2000}. 

PL occurs when a hole recombines with an electron, a process represented by a transition from the upper part of Fig.\ 4(b) to the lower part at identical $V_{g}$. Absorption is the opposite process, the creation of an electron-hole pair, represented by a transition from the lower part to the upper part of Fig.\ 4(b). Concerning the absorption process, for $V_g<V_3$ the absorption arises from a transition from the empty QD to X$^{0}$. At $V_{3}$, the absorption process changes to a transition from the one electron state to X$^{1-}$. Finally, absorption is blocked at $V_{4}$ when the electron ground state is doubly occupied, further occupation being prohibited through the Pauli principle. Furthermore, for $V_g<V_1$, the X$^0$ state lies above the hole-only state such that the final state after absorption is metastable through electron tunneling. This causes the absorption linewidth to increase. In all respects, the model reproduces the features in the experiment.

\begin{table}
\begin{ruledtabular}
\begin{tabular}{ll}
State of QD &Energy\\
\hline
empty & 0\\
e & $-E_{el}(V_{g})-E_{c}-E_{m}$\\
2e & $-2E_{el}(V_{g})-2E_{c}-4E_{m}+E^{ee}$\\
h & $E_{g}+E_{el}(V_{g})+E_{c}+E_{m}$\\
X$^{0}$ & $E_{g}-E^{eh}$\\
X$^{1-}$ & $E_{g}-E_{el}(V_{g})-E_{c}-E_{m}+E^{ee}-2E^{eh}$\\
\end{tabular}
\end{ruledtabular}
\caption{\label{tab:table1}The energies of the QD states in terms of the energies defined in Fig.\ 1 and $E_{el}(V_{g})=e(V_{s}+V_{g})/\lambda$, the electrostatic energy of charge $e$ at the site of the QD due to the applied and Schottky biases, $E_{m}$ the electrostatic energy given by the interaction of charge $e$ with its image charge $-e$ in the back contact, $E^{ee}$ the electron-electron on-site Coulomb energy, and $E^{eh}$ the electron-hole on-site Coulomb energy. The parameters were determined from the absorption data to be $E_c=134$ meV, $E^{ee}=23$ meV, $E^{eh}=29$ meV, $E_{g}=1306 $ meV, and $E_{m}=1.1$ meV. $\lambda=7$ from the geometry of the structure and $V_{s}$ was taken to be 0.62 V \cite{WarburtonPRB1998}.}
\end{table}

To make a quantitative analysis, and also to understand the behavior of the PL, we use the measured $V_{3}$, $V_{4}$ and the energy difference between the X$^{0}$ and X$^{1-}$ absorption lines to determine the parameters in the model, $E_{c}$, $E^{ee}$ and $E^{eh}$, by using the equations in Table \ref{tab:table1}. The parameters we find for the QD in Fig.\ 4 are listed in the caption to Table \ref{tab:table1}. We use these parameters to calculate the energies of the states in the upper part of  Fig.\ 4(b) each containing a hole and therefore to predict the behavior of the PL.

\begin{figure}
\includegraphics{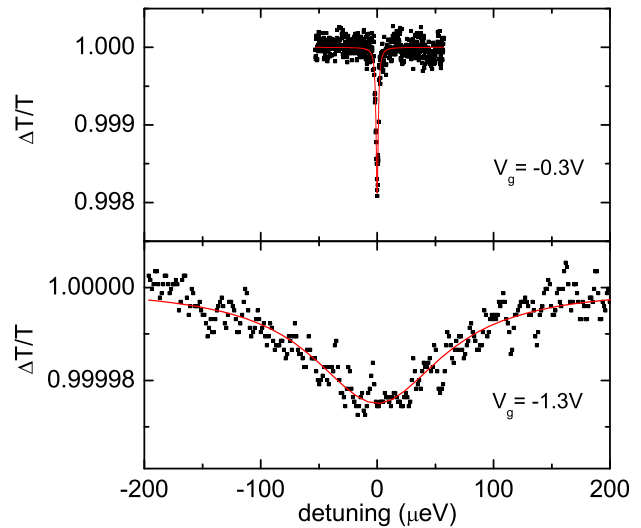}
\caption{Absorption in two different gate voltage regions. The upper graph shows the absorption resonance at a gate voltage of $-0.3$ V where the linewidth is not influenced by tunneling and the lower graph shows the resonance at $-1.3$ V where the linewidth is completely dominated by tunneling. The solid lines are Lorentzian fits to the data with linewidths 2.25 $\mu$eV in (a), 136 $\mu$eV in (b).}
\end{figure}

As explained above, the PL data in Fig.\ 4(a) have been corrected in gate voltage to eliminate the effect of the space charge which plays no role in the absorption measurements. A general consequence of the model, independent of the detailed set of parameters, is that after X$^{1-}$ radiative decay, the one electron state is only stable with respect to tunneling for $V_3<V_g<V_4$. This means that, in a similar way to the X$^{0}$ absorption at large negative biases, the X$^{1-}$ PL is broadened by an electron tunneling out within this voltage range. Despite the limited spectral resolution in the PL experiment, we observe experimentally a decrease in the X$^{1-}$ PL linewidth at a voltage just beyond the low-voltage edge of the X$^{1-}$ PL plateau and an increase close to the center of the X$^{1-}$ PL plateau. This is exactly as expected according to the model, and the voltage extent of the narrow linewidth region in the X$^{1-}$ PL plateau is exactly the voltage extent of the X$^{1-}$ plateau in the absorption experiment. We use this feature to correct for the space charge effects, determining $V_3$ in the PL experiment, and correcting it to equal $V_3$ in the absorption experiment. This allows a full quantitative understanding of the PL experiment.

Considering the initial states of the PL process, at large negative bias, the hole only state is favored which clearly cannot relax by emitting a photon explaining the quenching of the PL in this region. At $V_{1}$, the X$^{0}$ state is favored, allowing PL when the X$^{0}$ state decays to the empty state. At $V_{2}$, the X$^{1-}$ state is favored over the X$^{0}$, and the PL corresponds to a transition to the one electron state. As in the experiment, the model predicts that the X$^0$ to X$^{1-}$ transition at $V_2$ occurs at more negative bias than the X$^{0}$ to X$^{1-}$ transition in absorption at $V_3$. This is a generic feature, ultimately related to the fact that $E^{eh}$ is larger than $E^{ee}$, equivalently that the hole is more localized than the electron in a typical QD. The difference between $V_{2}$ and $V_{3}$ is directly proportional to the red shift between X$^{0}$ and $X^{1-}$. Quantitatively, we find excellent agreement with the PL experiment: our calculated voltages $V_{1}$ and $V_{2}$ agree to within the measurement uncertainty limited to 10 mV through the space charge effects. We note that the apparent small offset between the PL and the absorption X$^{0}$ energies at voltages between $V_1$ and $V_2$ is in all probability an artifact related to the relatively poor spectral resolution in the PL experiment. For $V_g>V_2$, the X$^{1-}$ initial state is favored in PL, and this persists over a large extent of $V_g$ due to a shell filling effect. The X$^{2-}$ forms only when the QD potential is reduced significantly for an electron to occupy the p shell \cite{WarburtonNature2000}. This explains why the X$^{1-}$ PL persists well beyond $V_4$.

\begin{figure}
\includegraphics{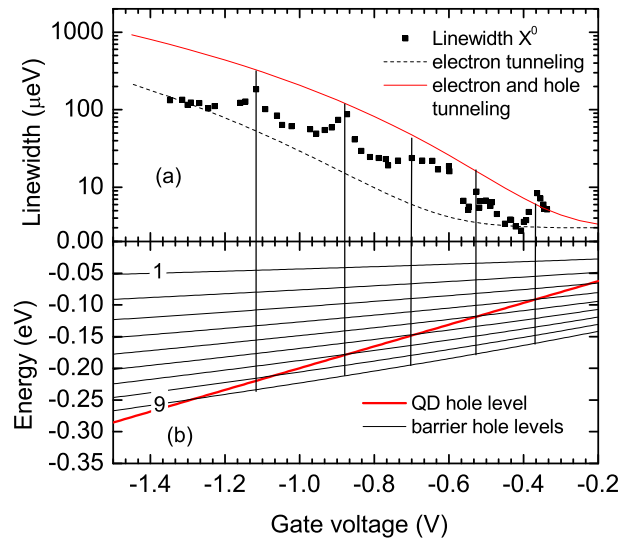}
\caption{Behavior of (a) the X$^{0}$ line width as function of gate voltage. The dashed line shows the expected linewidth calculated by WKB for electron tunneling and the solid line the linewidth for electron and hole tunneling. (b) Plot of the energy of the QD hole state and the energies of the 2D hole states confined at the interface between the capping layer and the blocking barrier versus gate voltage. The 2D states are labeled with an index with 1 describing the ground state. The energy zero lies at the top of the valence band at the interface between the capping layer and the blocking barrier.}
\end{figure}

The fact that the hole state is preferred over the X$^{0}$ state for very negative gate voltages is intimately related to the broadening of the absorption linewidth in the voltage region shown in Fig.\ 6(a). The broadening arises because the final state of the absorption has a reduced lifetime as the exciton cannot only decay radiatively but also tunnel out of the QD.  In the absorption experiment, both electron and hole tunneling must occur, as otherwise the remaining carrier would Coulomb-shift the QD out of resonance with the narrow band laser. This is consistent with photocurrent measurements where on similar structures, a photocurrent appears at the voltage where the PL quenches \cite{FindeisAPL2001,OultonPRB2002}. For the electron, the tunneling barrier between the QD and back contact decreases in extent with increasing negative gate voltage causing an increase in tunneling rate and concomitant increase in the absorption linewidth. We calculate electron and hole tunneling rates, $\gamma_{e}$ and $\gamma_{h}$, and from them a linewidth through $\Gamma=\Gamma_0+2\hbar \gamma_{e}+2\hbar \gamma_{h}$ where $\Gamma_0=3$ $\mu$eV and describes the voltage-independent linewidth. We use the WKB approximation with a QD height of 3 nm, an electron (hole) effective mass of $0.07 m_{0}$ \cite{FrickeEuPL1996} ($0.25 m_{0}$ \cite{WarburtonPRB1998}), electron ionization energy $E_c=134$ meV from the Coulomb blockade model and hole ionization energy $E_h=78$ meV determined from $E_c$, $E^{eh}$, the absorption energy and the GaAs fundamental band gap (1.519 eV). Although WKB is notoriously sensitive to the input parameters, the parameters are known well enough in this experiment to predict tunneling rates with less than an order of magnitude uncertainty. The result is plotted as a continuous line in Fig.\ 6(a). A comparison with the measured linewidth in Fig.\ 6(a) shows that the WKB model reproduces convincingly the trend in the experimental data, confirming that tunneling is the underlying mechanism for the increase in absorption linewidth. 

In addition to the exponential increase in absorption linewidth at large negative bias, oscillations in the linewidth are clearly visible (Fig.\ 6(a)). The calculations of the tunneling rate would suggest that at the resonances, WKB gives a very good account of the tunneling rates but that away from the resonances, hole tunneling is suppressed. We explain these features through a proper consideration of the hole tunneling out of the QD. The significant point is that hole tunneling occurs not into a 3D continuum, as for electron tunneling, but into a series of 2D states such that WKB is a reasonable approximation for the electrons but a poor approximation for the holes. The 2D hole states are defined by the triangular potential well at the interface between the capping layer and the blocking barrier. We calculate the energies of the hole 2D states as a function of gate voltage under the assumption of an infinite triangular barrier \cite{DaviesBook1998} and plot them in Fig.\ 6(b) together with the energy of the QD hole state. We find a remarkable correspondence between the experimental data and this calculation. Whenever the hole state is degenerate with a hole 2D state, there is a resonant increase in the absorption linewidth. This demonstrates that whenever there is a resonance, hole tunneling is much enhanced and is in fact faster than electron tunneling. Out of resonance, the hole tunneling is suppressed, and the linewidth correspondingly reduced. On resonance, the QD hole tunnels into a 2D state with in-plane momentum $k_{\parallel}$ close to zero. Out of resonance, tunneling can still proceed, but the hole must tunnel into a state with large $k_{\parallel}$. From the energy separation between the 2D hole levels in Fig.\ 6(b), we can estimate the largest $k_{\parallel}$ to be $\sim 5 \times 10^8$ m$^{-1}$. The lateral extent of the QD hole wave function is about $L_{\parallel}\sim 5$ nm, implying that tunneling is efficient up to wave vectors of about $1/L_{\parallel}=2.0 \times 10^8$ m$^{-1}$, smaller than that required away from a resonance. Hence, hole tunneling is suppressed out of resonance by the inability of the QD to provide the hole with sufficient in-plane momentum.

In conclusion, we have shown that the existence and the charge state of an exciton in a QD depends not only on the gate voltage but also on its creation process. For resonant exciton creation, which we detect with an absorption experiment, electron charging of the QD plays the main role whereas for non resonant exciton creation in a PL experiment, the hole is important, shifting the charging peaks to more negative voltages. We propose a model to explain these differences allowing us to calculate the Coulomb energies between holes and electrons in the QD. At large negative bias, the PL quenches and the absorption broadens as a consequence of electron and hole tunneling out of the QD. Electron tunneling increases monotonically with electric field but hole tunneling shows a series of oscillations in our structure as it is determined not by 0D-3D tunneling but by 0D-2D tunneling.

We acknowledge fruitful discussions with Atac Imamoglu and financial support from DFG (SFB 631), EPSRC, DAAD and the European Union Network of Excellence SANDiE.

\bibliography{paper}

\end{document}